\documentclass[12pt]{article}
 \usepackage{longtable}
\relax
\usepackage{amssymb}

\newcommand{\bo}{{\bar o}}

\def\bo{{\raise.15ex\hbox{\large$\Box$}}}               

\def\face{{\raise.2ex\hbox{$\displaystyle \bigodot$}\mskip-2.2mu \llap {$\ddot
        \smile$}}}                                      

\def\leftrightarrowfill{$\mathsurround=0pt \mathord\leftarrow \mkern-6mu
        \cleaders\hbox{$\mkern-2mu \mathord- \mkern-2mu$}\hfill
        \mkern-6mu \mathord\rightarrow$}       
\def\dvec#1{\vbox{\ialign{##\crcr
        \leftrightarrowfill\crcr\noalign{\kern-1pt\nointerlineskip}
        $\hfil\displaystyle{#1}\hfil$\crcr}}}           

\def\beq{\begin{equation}}
\def\eeq{\end{equation}}

\def\beqx{\begin{displaymath}}
\def\eeqx{\end{displaymath}}

\def\beql{\begin{eqnarray}}
\def\eeql{\end{eqnarray}}

\newcommand{\bea}{\begin{eqnarray}}
\newcommand{\eea}{\end{eqnarray}}

\def\[{\left [}
\def\]{\right ]}
\def\({\left (}
\def\){\right )}

\def\+{\oplus}

\begin{document}

\hbox{\hskip 12cm NIKHEF/2007-020  \hfil}
\hbox{\hskip 12cm IMAFF/FM-07/02  \hfil}

\vskip .5in

\begin{center}
{\Large \bf  Non-supersymmetric tachyon-free type-II and }

\vspace*{.2in}

{\Large \bf type-I closed strings from RCFT }

\vspace*{.7in}
{ B. Gato-Rivera}$^{a,b,}$\footnote{Also known as B. Gato}
{and A.N. Schellekens}$^{a,b,c}$
\\
\vskip .5in

${ }^a$ {\em NIKHEF Theory Group, Kruislaan 409, \\
1098 SJ Amsterdam, The Netherlands} \\

\vskip .2in

${ }^b$ {\em Instituto de Matem\'aticas y F\'\i sica Fundamental, CSIC, \\
Serrano 123, Madrid 28006, Spain} \\

\vskip .2in

${ }^c$ {\em IMAPP, Radboud Universiteit,  Nijmegen}

\end{center}

\begin{center}
\vspace*{0.5in}
{\bf Abstract}
\end{center}
We consider non-supersymmetric four-dimensional  closed string theories 
constructed out of tensor products of $N=2$ minimal models. Generically such theories
have closed string tachyons, but these may be removed either by choosing a non-trivial
partition function or a suitable Klein bottle projection. We find large numbers of examples
of both types.

\vskip 1in

\noindent
September 2007

\newpage


Supersymmetry is invaluable as a calculational tool in field theory and string theory,
but it is still not clear whether nature makes use of this virtue. Interesting
empirical evidence is the apparent convergence of the gauge couplings if gauginos 
and higgsinos are taken into account in their running.
Additional circumstantial evidence is the fact that low energy supersymmetry
automatically produces roughly the required amount of dark matter.
On the other hand, the usual argument that supersymmetry is ``required to stabilize
the gauge hierarchy" is in need of reassessment in view of the string theory landscape.
If the cosmological constant does not require any symmetry to ``remain" small, why would
there be such a requirement for the much smaller gauge hierarchy?

On the other hand, one might argue that supersymmetry is an essential ingredient of string theory. However,
the main ingredient of one-loop finiteness of closed strings is not
supersymmetry but modular invariance. Supersymmetry
is sufficient (but not necessary) to remove tachyons and the resulting divergence. 
It has been known since 1986  that one-loop finite, tachyon-free non-supersymmetric
heterotic strings can easily
be built in ten \cite{Dixon:1986iz}\cite{AlvarezGaume:1986jb}
as well as in four \cite{Kawai:1986ah}\cite{Lerche:1986cx} dimensions.
Despite these examples, there is a surprisingly widespread belief that absence
of tachyons in string theory requires supersymmetry. Another misconception we recently
learned about, and that finds its origin
in lack of familiarity with the early literature, is that string theory was once believed to predict
only negative cosmological constants, and that the positive observational result led to a 
re-consideration. In fact the issue of the cosmological constant in non-supersymmetric (and
of course tachyon-free) string theories was studied in several papers following the work
of G. Moore on ``Atkin-Lehner" symmetry \cite{Moore:1987ue}, and examples with both signs of the cosmological
constant were found, for example in 
\cite{Schellekens:1988az}.
Non-supersymmetric String Theories are usually divergent at higher loops
(see however \cite{Finite} for a set of exceptions and \cite{Harvey:1998rc} for
further discussion), but these divergencies can be
attributed to massless dilatons, which are unacceptable anyway. Of course we are
aware of the problems one encounters if one leaves the supersymmetry
highway at an earlier exit, but all of those problems have to be confronted at some
stage anyway, and the scenery may well be worth the price.

At present there is a huge literature on non-supersymmetric string
constructions and their implications, although this is still dwarfed by the amount
of work on supersymmetric constructions. An essential issue is the appearance of
tachyons, and many mechanisms have been proposed for getting rid of them  \cite{Finite}\cite{Non-tachyon}.
In this paper we will construct non-super- symmetric tachyon-free type-II and type-I closed strings
from tensor products of $N=2$ minimal models.
The models we consider are all obtained by tensoring $M$ minimal $N=2$ models
with a four-dimensional NSR sector, and imposing world-sheet supersymmetry.
The latter is, as always, done by extending the chiral algebra with all  pairs of
world-sheet superfields (``alignment currents")
of the $M+1$ factors. Note that since we wish to construct
fermionic strings, world-sheet supersymmetry is essential. However, this 
can also be achieved using $N=1$ building blocks. The reason we limit ourselves
to $N=2$ here is  simply that most of the necessary algorithms are already directly available
and optimized, because of  previous work in supersymmetric strings \cite{Dijkstra:2004cc}.

The resulting theory ${\cal G}$ has a chiral algebra consisting of the separate $N=2$
algebras of the building blocks, extended by the alignment currents. This theory is tachyonic.
On top of this we may add the following algebraic structures:

\begin{enumerate}
\item{A (further) extension ${\cal E}$ of the chiral algebra.}
\item{A modular invariant partition function (MIPF), denoted ${\cal M}$.}
\item{An orientifold choice, denoted ${\cal I}$.}
\end{enumerate}

Note that a MIPF can be of automorphism or extension type, or combinations thereof.
Even if it is purely of extension type, it still plays a different role than  ${\cal E}$.
The extension ${\cal E}$ defines the CFT used in the rest of the construction, in the sense
that only primaries, characters, O-planes (and eventually boundary states) will be used that respect
all of the symmetries of that extension. The extension in the MIPF projects out
closed string states that are non-local with respect to
it, but O-planes and boundary states that violate it are still permitted. Hence
the symmetry is  respected  in the bulk, but not necessarily by boundaries and crosscaps.
The introduction of orientifold planes is a first step towards type-I strings, of which we
only consider the closed sector here. These unoriented closed string theories will in general have tadpoles
due to closed string one-point functions on the crosscap. In some cases, these tadpoles may be cancelled
by introducing non-vanishing boundary state (Chan-Paton) multiplicities, but that possibility
will not be investigated here. Our main interest is to find out if it is possible to obtain a closed
string sector that is completely free of tachyons, and how often that occurs.

In principle, all three kinds of algebraic modifications listed above might remove the tachyons.
An obvious way to get rid of all the tachyons is to choose a space-time supersymmetric chiral
algebra extension ${\cal E}$. In general, there are many of these, but they can all be obtained 
as a basic extension by a spin-1 operator related to the space-time supercurrent, and further
extensions on top of that.  This is guaranteed to work, but will produce supersymmetric
strings and therefore not of interest here. Hence we will choose a non-supersymmetric ${\cal E}$.
This  leaves us with four distinct possibilities for removing the tachyons:

\begin{itemize}

\item{1. The action of the non-supersymmetric ${\cal E}$ itself.}
\item{2a. A MIPF including a space-time supersymmetric extension.}
\item{2b. A non-supersymmetric MIPF.}
\item{3. An orientifold choice.}
\end{itemize}

These four possibilities are the subject of our present investigation. Note that
tachyon-free models in class 2a are guaranteed to exist, but unlike the
supersymmetric ${\cal E}$ case, they do not lead automatically to supersymmetric
open strings, and have not been considered before in this context.
This case will enable us to study type-I models with  supersymmery in the bulk,
but not on the branes.

The total number of combinations ${\cal G}$, ${\cal E}$, ${\cal M}$ and ${\cal I}$ at
our  disposal is gigantic. The number of choices for ${\cal G}$ is 168 (and larger still
if we were to allow $N=1$ building blocks). The number of choices for ${\cal E}$ is typically
of order 100 to 1000, of which just one has been considered in \cite{Dijkstra:2004cc}, where a
search was done for standard-model-like supersymmetric open string spectra.
For each
${\cal E}$, the number of MIPFs ${\cal M}$ is of order 10 to 1000, and for each
${\cal M}$ the number of ${\cal I}$ is of order 10. We limit ourselves here to simple
current extensions, MIPFs and orientifolds, which are the only ones for which we have the
required calculational tools available \cite{Fuchs:2000cm}. Furthermore we limit ourselves 
to left-right symmetric MIPFs for the same reason. Hence the oriented theories we get
are of type IIB. 

In addition, some practical limits must be imposed to make this project manageable. In
\cite{Dijkstra:2004cc},
the supersymmetry extension ${\cal E}$ plays several r\^oles: it  automatically removes tachyons,
simplifies the tadpole conditions by combining NS and R tadpoles, drastically reduces
the number of primaries and boundary states to a reasonably sized set, and last
(and perhaps also least), it gives rise to a phenomenologically interesting supersymmetric
spectrum. Of these four features, the first, second and last are inevitably lost in the
present setting, but the third is essential if we ever want to explore standard-model-like
boundary state combinations. For example, the supersymmetric extension of the tensor
product $(3,3,3,3,3)$ (related to the ``Quintic") has 4000 primaries, but  without that extension it 
has 400000; for the tensor product $(6,6,6,6)$ these numbers are respectively 9632 and 2458624.
In the latter case the corresponding BCFT has 2458624 boundaries, for the charge conjugation
MIPF, out of which one has to choose three or four to  make a standard model configuration.
For this reason we consider here only extensions that yield at most
4000 primaries. Even at this stage, without considering boundary states,
a larger number is hard to deal with because the number of candidate tachyons
as well as the number of MIPFs tends to increase with the number of primaries.

Another issue that we have to deal with is that of permutation symmetry of identical $N=2$
factors in a given tensor product and outer automorphisms thereof.
Each $N=2$ factor has an outer automorphism, namely charge conjugation.
If these  symmetries  survive the extension ${\cal E}$, then
they act on the MIPFs and relate them to each other, as  $M \rightarrow P^{-1}MP$,
where $P$ is the symmetry and $M$ the modular matrix.  MIPFs related in this way give
rise to the same physics, and therefore should be identified. The supersymmetry
extension has the advantage of removing the outer automorphisms, because
the separate outer automorphisms of each factor act on the corresponding component
of the  supercurrent, changing it from $(S,\ldots,S,S,S,\ldots,S)$ to $(S,\ldots,S,S',S,\ldots,S)$. 
Here $S$ and  $S'$ are the two mutually conjugate Ramond simple currents of each factor,
with conformal weight $c/24$. These current combinations are always non-local
with respect to each other for any number of $S'$, except if all
components are flipped simultaneously. But the product of all outer automorphisms  
corresponds to charge conjugation in the full theory,  and  charge conjugation acts
trivially on the MIPF: $CMC=M$. This leaves the permutation symmetry, which is
never broken by the supersymmetric extension, and can be at most the permutation group of
9 elements for the tensor product $(1)^9$. This can still be handled, with some difficulty.
In the non-supersymmetric case  the situation is more complicated. First of all
we have to mod-out the action of all these symmetries on the extensions.
For the supercurrent this is easy: clearly $(S,\ldots,S,S,S,\ldots,S)$
is equivalent to all other choices obtained by replacing any $S$ by $S'$. Once
we have selected an extension we have to consider the surviving symmetry.
Now, the extension can break the automorphisms as well as the
permutation symmetry, but in the worst case none of them is broken, and we end up
with a symmetry group of $2^9 \times 9!$ elements for the tensor product $(1)^9$.

As was mentioned before, the total number of tensor combinations is 168. We have scanned all these
models for solutions of type 1, 2a, 2b and 3, and no solutions of type 1 were found. In  other words,
the only way to get rid of tachyons by means of a pure extension ${\cal E}$
is to choose the supersymmetric
extension.
Regarding solutions of the other three types,  we will not list all the ones of type 2a, since
their existence is essentially automatic: any simple current that is local with respect to 
the supercurrent $(S,\ldots,S,S,S,\ldots,S)$ can be used to generate a non-supersymmetric
extension ${\cal E}$ that allows a supersymmetric MIPF. Usually such currents exist, although
they do not always reduce the number of primaries below our practical upper limit of 4000.
The existence of solutions of type 2b and 3, on the other hand, is non-trivial. We present
our results for these cases in the table. The organization of this table is as follows.
Column 1 lists, in an obvious notation, the tensor product of the N=2 minimal models.
Column 2 lists the total number of extensions,
including the supersymmetric ones, and column 3 lists the number of extensions which
are non-supersymmetric and have 4000 or fewer primaries. These are the ones considered
in this paper. Column 4 lists the total number of MIPFs for the extensions considered. The
last two columns specify the number of tachyon-free models of oriented and unoriented type.
Obviously the oriented ones may still be orientifold projected, and since they are already
tachyon-free, all their orientifolds will be tachyon-free as well. Hence the total number
of closed sectors of type-I strings consists of the numbers in column 6, plus the ones
in column 5 multiplied with the number of orientifolds.

The reason that the last two columns contain four entries is as follows.
Apart from supersymmetric extensions of the chiral algebra or in the MIPF, one may also
encounter extensions that correspond to an embedding of the NSR fermions
in more than 4 dimensions. Note that in a type-IIB theory, N=4 space-time supersymmetry implies such an
extension to D=6, whereas N=8 implies an extension to D=10. In non-supersymmetric
type-IIB string theories, extensions to 6, 8 and 10 dimensions may occur, and in a few cases
even to 12 and 14 dimensions. In the table we have indicated this by giving in each field
in the last two columns four numbers, indicating respectively the number of cases 
with an NSR sector embedded in 4, 6, 8 and 10 dimensions. The larger dimensions occur
very rarely (D=12 for the tensor product $1^54^2$, and D=12 and D=14 for
$1^74$ and $1^9$). Note that the bosons $X^{\mu}$ describe a four-dimensional 
target space in all these cases.
We do not distinguish between higher dimensional embeddings for the
extensions ${\cal E}$ or the MIPFs in the table. Both occur, but only the latter are of interest for
future purposes, namely finding chiral open string spectra:  if the extension ${\cal E}$
is higher-dimensional all  spectra will be automatically non-chiral.

There is some small overcounting for a few of the free-field based models, namely tensor products
containing one or more factors $k=1$ or $k=2$. This happens because not all permutation symmetries
were taken into account for the tensor product $1^9$, and also because in a few other cases
there are some degeneracies
that are particular  for free-field models and that  have not been taken into account. The table
only contains tensor products for which non-supersymmetric tachyon-free spectra were found 
corresponding to solutions of type 2b and 3.
Tensor products with high values of the $N=2$ minimal model parameter $k$ are absent mainly because
their extensions violate our upper limit of 4000 on the number of primaries.

In conclusion: 
we have compiled a large database of non-supersymmetric tachyon-free type-IIB and type-I
closed strings.  We think that the existence of these theories is of interest in its own right. Furthermore,
in the future these theories will serve as a starting point to study non-supersymmetric tachyon-free 
open string theories, as has been done for the supersymmetric open string models in \cite{Dijkstra:2004cc} 
and \cite{Anastasopoulos:2006da}. The hope would be to find non-supersymmetric realizations
of the standard model spectrum. This is still a huge challenge, because in addition to
the closed string tachyons, we will have to deal with open string tachyons as well as
separate Neveu-Schwarz and Ramond tadpoles. However, the fact that the non-supersymmetric, non-tachyonic closed 
string database
is so huge may give us a chance to achieve that goal.

\vskip .2in
\noindent
{\bf Acknowledgements:}
\vskip .2in
\noindent
We thank Elias Kiritsis and Florian Gmeiner for useful conversations. This work has been partially 
supported by funding of the spanish Ministerio de Educaci\'on y Ciencia, Research Project
FPA2005-05046. The
work of A.N.S. has been performed as part of the programs
FP 52  and FP 57 of Dutch Foundation for Fundamental Research of Matter (FOM). 



{
\small
\LTcapwidth=14truecm
\begin{center}
\begin{longtable}{|l|l|l|l|l|l|}\caption{\em Summary of tachyon-free solutions of type 2b and 3}\label{tbl:TableOne}\\
 \hline \multicolumn{1}{|l|}{Tensor}
& \multicolumn{1}{l|}{Ext.}
& \multicolumn{1}{l|}{$<= 4000$}
& \multicolumn{1}{l|}{MIPFs}
& \multicolumn{1}{l|}{Oriented}
& \multicolumn{1}{l|}{Unoriented}\\ \hline
\endfirsthead
\multicolumn{6}{c}%
{{\bfseries \tablename\ \thetable{} {\rm-- continued from previous page}}} \\
\hline \multicolumn{1}{|c|}{Tensor}
& \multicolumn{1}{l|}{Ext.}
& \multicolumn{1}{l|}{$<= 4000$}
& \multicolumn{1}{l|}{MIPFs}
& \multicolumn{1}{l|}{Oriented}
& \multicolumn{1}{l|}{Unoriented}\\ \hline
\endhead
\hline \multicolumn{6}{|r|}{{Continued on next page}} \\ \hline
\endfoot
\hline \hline
\endlastfoot
(1,10,22,22) & 303 & 19 & 158 & 0,0,0,0 & 1,0,0,0 \\
(2,6,8,38) & 538 & 68 & 2384 & 0,0,0,0 & 4,0,0,0 \\
(2,6,10,22) & 1046 & 142 & 2982 & 0,0,0,0 & 2,0,0,0 \\
(2,6,14,14) & 733 & 158 & 5064 & 0,0,0,0 & 19,0,0,0 \\
(2,10,10,10) & 261 & 48 & 2088 & 0,0,0,0 & 6,0,0,0 \\
(3,6,6,18) & 88 & 16 & 188 & 0,0,0,0 & 1,0,0,0 \\
(4,4,6,22) & 398 & 46 & 1512 & 0,0,0,0 & 9,0,0,0 \\
(4,4,8,13) & 70 & 8 & 90 & 0,0,0,0 & 2,0,0,0 \\
(4,4,10,10) & 319 & 62 & 3238 & 0,0,0,0 & 79,0,0,0 \\
(4,6,6,10) & 378 & 95 & 3410 & 0,0,0,0 & 48,0,0,0 \\
(6,6,6,6) & 191 & 80 & 5254 & 74,0,0,0 &  238,0,0,0 \\
(1,1,4,6,22) & 316 & 85 & 3218 & 0,0,0,0 & 1,0,0,0 \\
(1,1,4,7,16) & 191 & 57 & 946 & 0,0,0,0 & 2,0,0,0 \\
(1,1,4,10,10) & 222 & 54 & 1748 & 0,0,0,0 & 12,0,0,0 \\
(1,1,6,6,10) & 88 & 26 & 1348 & 0,0,0,0 & 31,0,0,0 \\
(1,2,2,7,16) & 147 & 46 & 1376 & 0,0,0,0 & 4,0,0,0 \\
(1,2,2,10,10) & 341 & 126 & 10180 & 22,4,0,0 & 198,0,0,0 \\
(1,2,2,6,22) & 768 & 245 & 17468 & 20,0,0,0 & 73,0,0,0 \\
(1,2,4,4,10) & 463 & 188 & 26508 & 4,0,0,0 & 340,0,0,0 \\
(1,2,4,6,6) & 374 & 178 & 24364 & 20,0,0,0 & 370,26,0,0 \\
(1,4,4,4,4) & 192 & 74 & 5292 & 68,0,0,0 & 241,14,0,0 \\
(2,2,2,3,18) & 216 & 88 & 6092 & 130,66,0,0 & 0,0,0,0 \\
(2,2,2,4,10) & 1133 & 557 & 223978 & 2264,520,0,0 & 6334,784,0,0 \\
(2,2,2,6,6) & 1155 & 644 & 271198 & 1808,356,0,0 & 8988,1256,0,0  \\
(2,2,3,3,8) & 63 & 26 & 816 & 0,0,0,0 & 4,0,0,0 \\
(2,2,4,4,4) & 333 & 130 & 33804 & 72,48,0,0 & 635,40,0,0 \\
(3,3,3,3,3) & 12 & 3 & 14 & 0,0,0,0 & 1,0,0,0 \\
(1,1,1,1,5,40) & 36 & 10 & 162 & 0,12,0,0 & 0,0,0,0 \\
(1,1,1,1,7,16) & 123 & 61 & 1160 & 15,16,0,0 & 0,0,0,0 \\
(1,1,1,1,8,13) & 36 & 12 & 186 & 0,6,0,0  & 0,0,0,0 \\
(1,1,1,1,10,10) & 78 & 29 & 1208 & 16,24,0,0 & 1,1,0,0 \\
(1,1,1,1,6,22) & 108 & 35 & 892 & 0,8,0,0 & 0,0,0,0 \\
(1,1,1,2,4,10) & 228 & 106 & 8888 & 16,24,0,0 & 39,3,0,0 \\
(1,1,1,2,6,6) & 88 & 43 & 3652 & 0,0,0,0 & 0,16,0,0 \\
(1,1,1,4,4,4) & 197 & 113 & 8534 & 430,95,0,0 & 395,78,0,0 \\
(1,1,2,2,2,10) & 216 & 100 & 16972 & 408,148,0,0  & 676,0,0,0  \\
(1,1,2,2,4,4) & 265 & 164 & 49008 & 160,120,0,0 & 396,172,0,0 \\
(1,2,2,2,2,4) & 546 & 403 & 388155 & 2912,1583,0,387 & 4180,1564,0,0 \\
(2,2,2,2,2,2) & 754 & 617 & 2112682 & 17680,12560,0,1942 & 105653,43836,6818,4202 \\
(1,1,1,1,1,2,10) & 56 & 31 & 2984 & 28,52,0,0 & 0,0,0,0 \\
(1,1,1,1,1,4,4) & 120 & 80 & 8668 & 270,200,26,0 & 97,86,0,0 \\
(1,1,1,1,2,2,4) & 126 & 82 & 12832 & 0,84,32,0 & 27,50,4,0 \\
(1,1,1,2,2,2,2) & 120 & 91 & 38228 & 0,448,0,186 & 0,416,0,0 \\
(1,1,1,1,1,1,1,4) & 60 & 41 & 4426 & 218,190,95,0 & 9,11,8,0 \\
(1,1,1,1,1,1,2,2) & 35 & 24 & 2838 & 0,18,24,0 & 0,0,0,0 \\
(1,1,1,1,1,1,1,1,1) & 289 & 202 & 161774 & 52058,17568,5359,0 & 41168,10292,3993,478 \\
\end{longtable}
\end{center}
}


\newpage

\bibliography{REFS}
\bibliographystyle{lennaert}

\end{document}